# Epsilon-Near-Zero Al-Doped ZnO for Ultrafast Switching at Telecom Wavelengths: Outpacing the Traditional Amplitude-Bandwidth Trade-Off

N. Kinsey,[1] C. DeVault,[2] J. Kim,[1] M. Ferrera,[1,3] V.M. Shalaev,[1] and A. Boltasseva[1,*]

[1] School of Electrical & Computer Engineering and Birck Nanotechnology Center, Purdue University, West Lafayette IN, 47907, USA
[2] Department of Physics and Birck Nanotechnology Center, Purdue University, West Lafayette IN, 47907, USA
[3] School of Engineering and Physical Sciences, Heriot-Watt University, Edinburgh, Scotland EH14 4AS, UK
*Corresponding author: aeb@purdue.edu

Transparent conducting oxides have recently gained great attention as CMOS-compatible materials for applications in nanophotonics due to their low optical loss, metal-like behavior, versatile/tailorable optical properties, and established fabrication procedures. In particular, aluminum doped zinc oxide (AZO) is very attractive because its dielectric permittivity can be engineered over a broad range in the near infrared and infrared. However, despite all these beneficial features, the slow (> 100 ps) electron-hole recombination time typical of these compounds still represents a fundamental limitation impeding ultrafast optical modulation. Here we report the first epsilon-near-zero AZO thin films which simultaneously exhibit ultra-fast carrier dynamics (excitation and recombination time below 1 ps) and an outstanding reflectance modulation up to 40% for very low pump fluence levels (< 4 mJ/cm$^2$) at the telecom wavelength of 1.3 μm. The unique properties of the demonstrated AZO thin films are the result of a low-temperature fabrication procedure promoting oxygen vacancies and an ultra-high carrier concentration. As a proof-of-concept, an all-optical AZO-based plasmonic modulator achieving 3 dB modulation in 7.5 μm and operating at THz frequencies is numerically demonstrated. Our results overcome the traditional "modulation depth vs. speed" trade-off by at least an order of magnitude, placing AZO among the most promising compounds for tunable/switchable nanophotonics.

## 1. Introduction

### i. TCO Based Tunable Nanophotonics

With the advent of plasmonics and metamaterials, it has become of paramount importance to develop new CMOS-compatible platforms possessing metal-like behavior and great optical transparency. Transparent conducting oxides (TCOs) are a class of materials which are well known for their optical transparency combined with high electrical conductivity, and consequently, they appear to be critical for enabling novel tunable nanophotonic devices. TCOs such as indium doped tin oxide (ITO) and indium+gallium doped zinc oxide (IGZO) have been widely used in industrial applications for touchscreen displays, transparent electrodes on solar panels, and more recently in televisions [1, 2]. As a result of their widespread industrial use, many CMOS compatible deposition techniques have been demonstrated which enable the growth of high quality, room temperature thin films [3-6]. One of the valuable assets of TCOs is the intrinsic tunability of the optical properties. Since TCOs can sustain extremely heavy doping without a consistent degeneration of their morphological structure, very high carrier concentrations (~$10^{20}$ cm$^{-3}$) can be achieved [7-11]. The optical properties of these materials can be tailored in many ways such as by altering the deposition conditions, through post‑processing steps (e.g. annealing) and/or by varying the material stoichiometry [8, 10, 12-16]. This enables one material to serve several roles. For instance, a lowly doped TCO can be used as a dielectric material in the near infrared (NIR) range with an index near that of glass, while a highly-doped TCO can serve as a metal to support surface plasmon oscillations [17, 18].

   Additionally, TCOs exhibit the potential for both electrical and optical excitation allowing a dynamic modification of their optical properties [11, 19-22]. The application of an electric field results in carrier accumulation or depletion causing a temporary change in the optical properties within a thin layer (~2-5 nm) [23]. Even though this method can have extremely low energy dissipation in the fJ/bit range [24], the overall effect is generally limited by the RC‑delay of the specific electrical system to the GHz range in practice, although theoretically faster speeds could be attained.

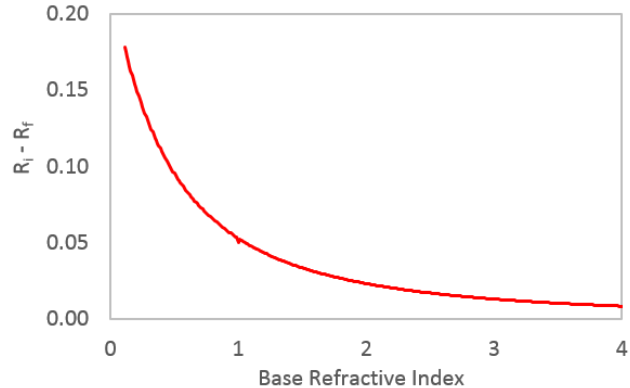

Fig. 1. Absolute change in the reflection of a purely real (i.e. Im{n}= 0) material versus base refractive index provided an index change of $\Delta n$ = -0.1. The magnitude of reflection is calculated using the Fresnel equations at a single interface between air ($n_o$ = 1) and the material assuming normal incidence. Note that operating in the epsilon-near-zero regime (i.e. n < 1) provides larger absolute changes in the reflection for the same change in the refractive index.

Optical excitation of TCOs is also a potential method for dynamically tuning the optical properties of the film similar to the approach demonstrated with other materials such as semiconductors and recently ITO nanorods [21, 25, 26]. In this approach, light with energy greater than the band-gap is used to excite valence band electrons into the conduction band, which in turn alters the properties of the material until they recombine and the material returns to equilibrium. This method has the benefit of generating free carriers throughout the bulk of the entire film (~100's nm thick) instead of only a few nanometers and, in principle, is only limited by the material absorption time (~10 fs) and recombination time. In particular, a slow electron-hole recombination time is a fundamental burden for the realization of numerous all-optical devices such as ultrafast switches, modulators, multiplexers, etc. Our study is directly focused on solving this issue as well as increasing the limited available information about the opto-electric dynamics of TCOs. Consequently, the remainder of this work will is dedicated to the study of photo-induced effects and how to overcome the typical trade-off between speed and amplitude.

**ii. Speed vs Amplitude Trade-Off**
Commonly, in semiconductors (Si, GaAs, etc), the recombination time is on the order of 100's of picoseconds to nanoseconds, but can be drastically enhanced through surface recombination, nanoparticle trapping, and defect recombination centers [27-29]. Low temperature (< 400°C) fabrication processes can be engineered in order to achieve the desired ultrafast recombination, but come at the cost of a remarkable deterioration of fundamental optical (e.g. higher losses) and electronic (e.g. lower carrier mobility) properties [29]. In addition to this, the relative change in both the real and the imaginary part of the material's permittivity still remains very small (a few percent) and unsuitable for numerous fundamental applications (modulators, detectors, etc.).

To enhance the amplitude of the material response, we propose to operate in the epsilon near zero (ENZ) regime [22, 30-32]. As shown in Fig. 1, operating in the ENZ regime (i.e. n < 1) produces larger absolute changes in the reflection for a fixed change in the refractive index ($\Delta n$ = -0.1 in a purely real medium). This can be understood by considering the change in magnitude with respect to the initial index e.g. a change of -0.1 for an initial index of 0.2 is 50%, where for an initial index of 2 it is only a 5% change. Consequently, this makes the ENZ regime an attractive region for maximizing the performance of dynamic devices.

Traditional semiconductors cannot achieve ENZ operation at the technologically important telecommunication wavelengths due to their large background permittivity and low dopant solubility - even other TCOs struggle with this problem [33]. While some TCOs can achieve ENZ properties at telecom, all currently published works have yet demonstrate a large optical response and a short recombination time.

In this work, we develop a CMOS-compatible, oxygen-deprived aluminum doped zinc oxide (AZO) film which achieves both of the desired properties simultaneously. Our unique material recovers in less than 1 ps with a relative variation of the transient reflectivity ($\Delta R/R_o$) and transmissivity ($\Delta T/T_o$) as large as 40% and 30%, respectively (roughly one order of magnitude larger than previous ultrafast semiconductor based films [28]). Both the ultrafast recombination and the extremely high intrinsic carrier concentration (~$10^{21}$ cm$^{-3}$) are a direct consequence of our unique low-temperature fabrication process, which induces oxygen vacancies to produce the required number of donor centers (see Results and Supplementary Information).

Consequently, our AZO thin films satisfy key attributes of dynamic materials simultaneously: 1) Large intrinsic carrier concentration to enable ENZ properties at the technologically important telecommunication wavelengths (which enhances the optical/electrical tunability); 2) Low-temperature deposition which does not impede conductivity and transparency; 3) Maturity of the fabrication and nano-patterning process; 4) CMOS-compatibility; and 5) Ultrafast electron-hole recombination dynamics. In order to further investigate the applicability of our oxygen-deprived AZO, we also propose a scheme for an all‑optical plasmonic modulator using CMOS-compatible materials which achieves 3 dB modulation in 7.5 μm with less than 0.1 dB/μm insertion loss. As a result, our work overcomes the traditional "modulation depth vs. speed" trade-off by at least an order of magnitude, placing our AZO films among the most promising compounds for tunable/switchable nanophotonics.

## 2. Results

### i. Oxygen-Deprived AZO Thin Films

A 350 nm thick AZO film (hereto referred to as-grown AZO) was deposited at room temperature on a fused silica substrate by pulsed laser ablation of a 2% wt. aluminum doped zinc oxide target under low oxygen partial pressure to induce oxygen vacancies (for more information see the Supplementary Information) [13]. These oxygen vacancies push the already high carrier concentration provided by the doping process even further, giving rise to an extremely high intrinsic carrier concentration within the material that can be controlled by varying the oxygen pressure during deposition, or through annealing in an oxygen rich environment [12]. To dynamically modify the carrier concentration within the film, the material must be pumped by light with photon energy greater than the bandgap (~3.5 eV). The excess free electrons then cause a change in the optical properties (reflection, transmission, and absorption) of the probe through the Drude dispersion, which is added to the background or intrinsic permittivity of the material. Then, the system recovers through the recombination of excess carriers, returning to its original state.

### ii. Preliminary Material Characterization

To determine the proper pump and probe wavelengths, the linear optical properties of the as-grown AZO film are required (see Fig. 2). The permittivity of the sample was extracted from spectroscopic ellipsometry using a Drude + Lorentz model (see Supplementary Information) which describes the intrinsic material response. Consequently, the ENZ regime of the material (1.1 - 1.5 μm) is extracted and shaded red in Fig. 2(a). Additionally, the linear transmission spectrum of the sample is shown in Fig. 2(b), which illustrates the band edge of the material. As can be seen from the both curves, the band edge of the as-grown AZO film lies at ~350 nm (shaded green). Consequently, light shorter than 350 nm must be used to pump the AZO. Therefore, a combination of 1.3 μm and 325 nm was chosen as it provides a pump greater than the bandgap energy and a probe at a technologically relevant wavelength that is also near the ENZ point of the material (see Fig. 2(a)). While more efficient absorption of the pump could be achieved with slightly shorter wavelengths, it is beneficial to keep the pump wavelength at as low an energy as possible to aid the efficiency of conversion as well as to minimize losses in optics such as lenses, mirrors, etc.

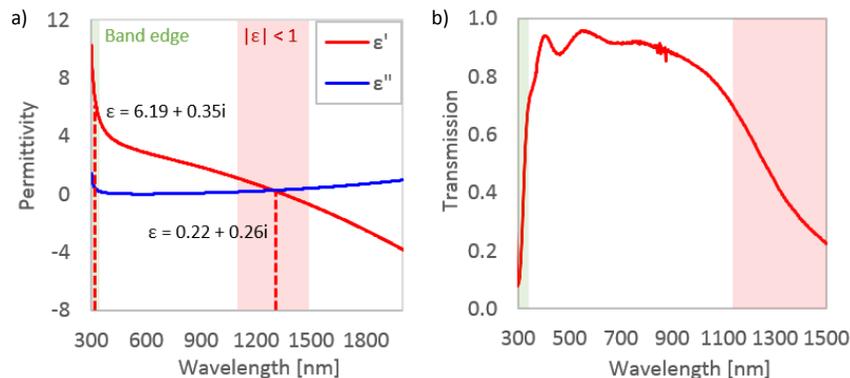

Fig. 2. a) Complex permittivity of the 350 nm as-grown AZO film as extracted from spectroscopic ellipsometry. The green shaded area represents wavelengths above the band edge of the AZO and the red shaded area represents the ENZ regime i.e. $|\varepsilon| < 1$. The permittivity of the AZO at the two test wavelengths are listed for reference. b) Transmission spectrum of the 350 nm AZO film obtained using a spectrophotometer.

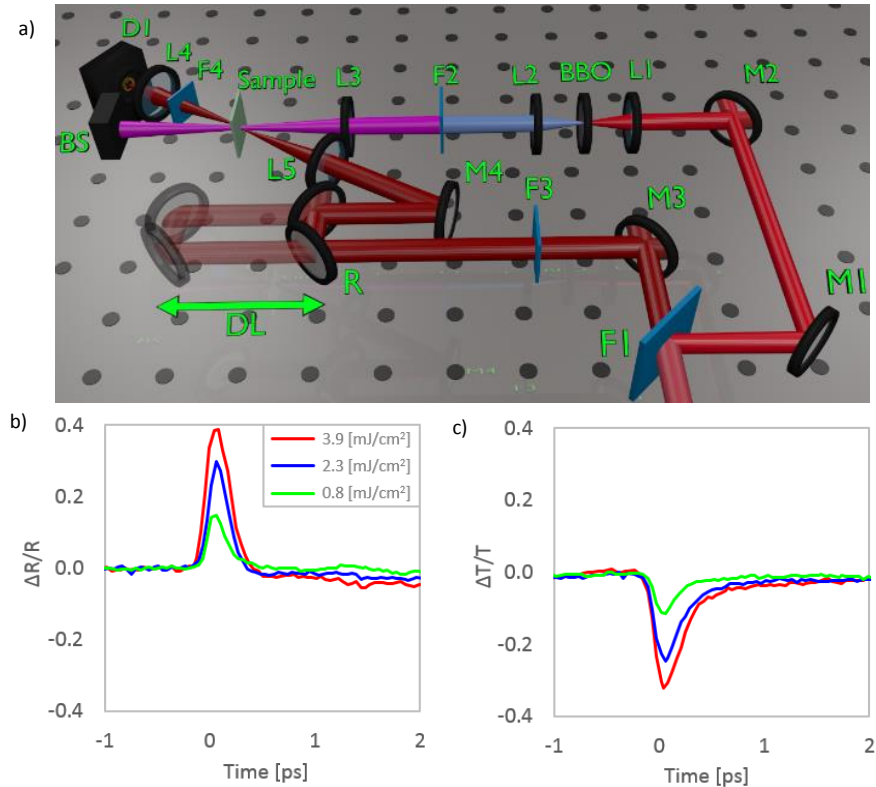

Fig. 3. a) Schematic of the pump-probe setup. Filter F1 transmits 1.3 µm and reflects 650 nm light. F3 removes residual 650 nm. R provides a delay line. Lenses L1 and L2 focus the light onto BBO to generate 325 nm light. Filter F2 removes residual 650 nm. Lenses L3 and L5 focus light onto the sample. F4 filters and stray light for detector D1. Normalized change in the b) reflected power and c) transmitted power as a function of the delay time between the pump and probe pulses.

### iii. Optical Setup
An illustration of the pump-probe system used to test the as-grown AZO films is depicted in Fig. 3(a). Collinear 1.3 µm and 650 nm light is brought into the system from the bottom right and separated by a dichroic filter (F1). 650 nm is focused onto a BBO crystal to generate the second harmonic, and filtered to remove any residual 650 nm. 1.3 µm is filtered by a long-pass filter (F3) to remove any 650 nm light and directed through a computer controlled retroreflector providing a delay line. Both 1.3 µm and 650 nm beams are focused onto the sample where the 1.3 µm is filtered and collected by an InGaAs detector D1. A more detailed description of the experimental methods are found in the Supplementary Information.

### iv. Transient Reflectivity and Transmissivitiy
The normalized reflection and transmission as a function of the pump-probe pulse delay is shown in Fig. 3(b), (c), respectively, for the 350 nm as-grown AZO thin film under several incident pump fluence levels. Upon closer examination of these results, a few important points arise, 1) The transient effect is fully recovered (from start to finish) within 1 ps (greater than 1 THz potential modulation speed) with a relaxation time less than 500 fs; 2) The change in the transmission/reflection is large, with all measurements having at least a 10% variation and a peak change of 40%; 3) The UV fluence used is at least an order of magnitude below the typical damage threshold for dielectrics (~100 mJ/cm$^2$ for femtosecond pulses [34]) indicating that the effect could be further increased; 4) A total pulse energy of only a few micro Joules is required to induce the effects shown here; 5) The material thickness is compatible with current integrated photonic technologies which is especially critical since the material properties (e.g. carrier concentration) may differ drastically in thin films when compared to bulk films.

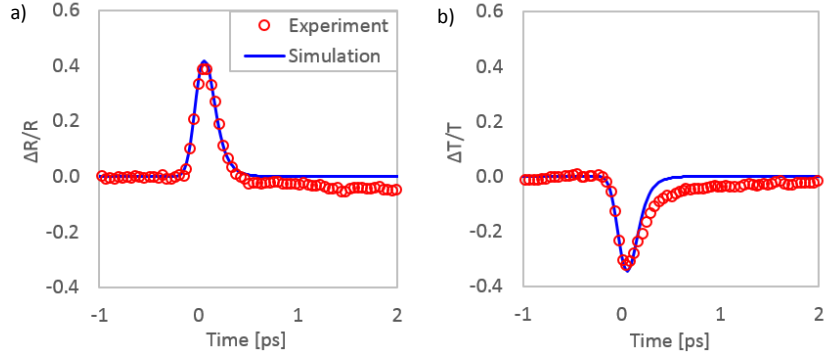

Fig. 4. Fitted change in a) reflection and b) transmission for the 350 nm thick as-grown AZO film under a fluence of 3.9 mJ/cm$^2$.

Table 1. Summary of the extracted properties of the 350 nm as-grown AZO sample including the intrinsic concentration. The max. ΔR and ΔT values are the peak relative change in reflection and transmission, respectively, and Δn + Δk correspond to the maximum change induced in the material properties.

| Pump Fluence [mJ/cm$^2$] | Max. ΔR [%] | Max. ΔT [%] | Intrinsic Concentration [cm$^{-3}$] | Rec. Time [fs] | Avg. Induced Carriers [cm$^{-3}$] | AZO Δn + iΔk |
|---|---|---|---|---|---|---|
| 1.0 | 15 | -12 | 9.8 × 10$^{20}$ | 100 | 0.2 × 10$^{20}$ | -0.07 + i0.07 |
| 2.4 | 30 | -25 | 9.8 × 10$^{20}$ | 92 | 0.5 × 10$^{20}$ | -0.14 + i0.16 |
| 3.6 | 39 | -32 | 9.8 × 10$^{20}$ | 88 | 0.7 × 10$^{20}$ | -0.17 + i0.25 |

One of the key parameters for dynamic materials is the maximum achievable change in the optical properties, which in this case is the result of excess carrier generation. Thus, it is critical to understand the carrier concentration that was achieved during the experiment. To extract this value, a theoretical model has been used which is then fitted with the experimental results (see Supplementary Information). In Fig. 4 the results of the fitting are shown, where an excellent agreement is found between calculation and experiment. However, there is a slight discrepancy in the transmission measurement towards the end of the decay (a small offset, ~3%) in both the reflection and transmission (i.e. they do not return to zero). This small offset is believed to be the result of lattice heating from excited electrons relaxing to the bandedge, which plays a role significantly longer timescales (100's picoseconds to nanoseconds) [35]. This minimal alteration of the optical properties is not permanent (e.g. material damage) and the AZO returns to equilibrium within one millisecond (repetition rate of the laser source). In addition, these effects have very little influence upon the ultrashort response of the as-grown AZO and therefore do not significantly influence our extraction of the excess carrier concentration or decay time in the sample.

With a successful fitting achieved, the carrier distribution can be obtained. From the model, an average excess carrier density of $0.7 \times 10^{20}$ cm$^{-3}$ is estimated within the sample's thickness of 350 nm, corresponding to a peak of $1.4 \times 10^{20}$ cm$^{-3}$ near the surface at time t = 0, which is similar to the values reported for optically modified ITO nanorods [21]. However, for electrical tuning a peak $\delta N = 6.7 \times 10^{20}$ cm$^{-3}$ has been achieved, although it is important to remember that these densities are achieved only within a few nanometers of the surface and not throughout the bulk of the material [23]. In fact, considering an accumulation layer thickness of 5 nm for the δN listed above, the raw number of carriers induced using optical excitation is 10× more than what was reportedly achieved through electrical biasing.

The excess carrier density and recombination time have been extracted for the listed fluences and are presented in Table 1 together with the maximum relative change in reflectivity and transmissivity, intrinsic carrier concentration, electron-hole recombination time, and the variation in the complex refractive index. Here, we see that the range of average carrier densities induced in the sample is on the order of $10^{19}$ cm$^{-3}$ (roughly 5-10% of the intrinsic concentration) being measured for an incident energy of a few micro-Joules. This is also particularly interesting considering the large change in the optical properties which is observed, and likely arises from operating in the ENZ regime.

## 3. Discussion

It is important to mention that the carrier densities reported in Table 1 underestimate the actual carrier density achieved within the material after pumping due to the temporal resolution limit of the system. Although the actual induced carrier density is larger, we report in Table 1 a conservative estimation of the achieved carrier density averaged over the thickness of the material as extracted from the experiment using two finite width pulses.

The ultrafast temporal response is believed to be the result of defect enhanced Shockley‑Read‑Hall recombination which arises due to the way the AZO is grown [36]. By depositing the AZO under a severe lack of oxygen, a large density of oxygen vacancies are created in the film which serve two purposes: to provide extra intrinsic carriers, and generate deep level defects.[37, 38] These deep level defects dramatically reduce the recombination time for electrons resulting in ultrafast recombination. Similar situations have been investigated in low temperature deposited gallium arsenide (LT-GaAs), as well as in several other specially grown materials, where the timescale of the process is attributed to defect enhanced recombination [27, 29]. However, one key distinction of as-grown AZO from other materials is the magnitude of the effect. Generally for LT-GaAs and others, the measured change in refractive index is < 2%, still abiding by the amplitude‑bandwidth trade-off. Our as-grown AZO is shown to have an order of magnitude larger response, which as we mentioned in previous sections is largely the result of operating in the ENZ regime, than previous works while achieving speeds that are generally 2-3× faster [29]. This speed can also be compared to the optical pumping of traditional semiconductors such as silicon, where typical recombination times are on the order of 0.1 – 1 ns, limiting the overall speed to 1 – 10 GHz (three to four orders of magnitude slower than the as-grown AZO film) [25]. Electrical tuning methods in silicon, which make use of carrier accumulation/depletion, rely on diffusion to remove carriers and can be faster. However, these devices are well known to be limited by the capacitor charging time (RC delay) to a few 10's of GHz and maybe up to 100 GHz – still at best an order of magnitude slower than the as-grown AZO [39-41]. As for recent works optically tuning ITO nanorods, we note at a maximum effect 2× larger and 6× faster for lower pump fluence levels than was previously reported, without the need for complicated nanorod structures [21].

Furthermore, the fact that the total carrier density is not radically changing to produce the modulation effect results in recombination times which are relatively constant. This can be an important feature for practical devices such that the temporal response of the system does not change based on pump fluence, enabling the use of the minimal energy without sacrificing speed, or for more specific applications such as high-speed analog encoding and data.

In addition, there is the potential for an inverted modulator by properly controlling the initial permittivity of the as-grown AZO film. Here, the base permittivity of the film is already near zero (i.e. less than unity) such that inducing additional free electrons causes the material to appear "more-metallic", thereby increasing the reflection. However, if the as-grown AZO film was grown with a lower intrinsic carrier concentration such that the material's permittivity is greater than unity at 1.3 μm, adding additional carriers would cause a decrease in the reflection. This flexibility enables the potential for both positive and negative logic devices from the same material by simply altering the deposition conditions.

Finally, the ultrafast response of the as-grown AZO opens many additionally applications beyond just high-speed data processing. One application could be in the field of THz wave generation – consistently a challenge for the field. By pumping the as-grown AZO with an ultrafast 325 nm pulse, a THz modulation of the conductivity is achieved which, when combined with the proper electrical circuitry, can serve as a THz radiation source. The quick response is also valuable for measuring the temporal shape of short optical pulses, which can be quite difficult in the ultraviolet range due to the inefficiency of traditional materials such as silicon, and GaAs, assuming they were capable of such an ultrafast response. Combining an as-grown AZO based detector with built-in electrical delays and phase locking electronics, an integrated opto-electric sampling detector can be realized which is capable of a few picosecond resolution. This technique could be much more compact and simple to use than traditional autocorrelators requiring two optical beams.

## 4. Integrated All-Optical Plasmonic Modulator

As a demonstration of the applicability and versatility of our as-grown AZO films, a simple plasmonic modulator is proposed which enables all-optical control of the signal amplitude at ultrafast rates. All optical devices promise extremely fast operational speeds, well beyond what is currently attainable with electronics or electro-optic hybrid devices. This dramatic increase in bandwidth is crucial for continuing the exponential improvement of computational systems as well as for next-generation communication networks. Many different materials and methods have been suggested to generate the desired temporal dependence such as material phase/structural changes [42-44], thermos−optic effects [45], phase shifting through carrier injection [39, 46], nonlinearities [47, 48], and electro−absorption [20, 23, 49, 50]. While some of these devices have been shown to achieve operational speeds greater than 10 GHz with appreciable modulation, for many modulators there is still a trade‑off between the speed and modulation depth [51]. More recently, transparent conducting oxides (TCOs) have received attention for their tunability, leading to several devices with some of the highest reported performance [11, 21, 22, 52, 53]. For more information on TCO based modulators see the following reviews and references therein [51, 54].

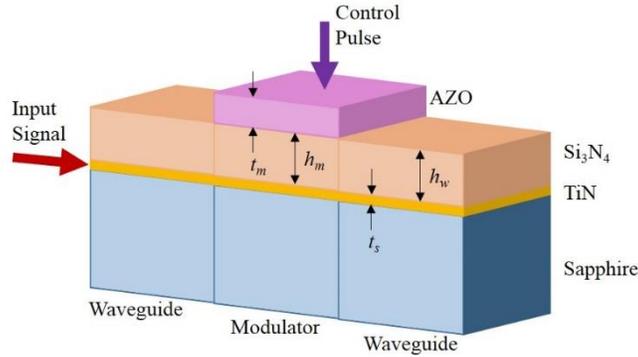

Fig. 5. Schematic of the as-grown AZO plasmonic modulator. The height of the silicon nitride cladding in the waveguide ($h_w$) was optimized using eigenmode analysis and set at 285 nm. The thickness of the TiN strip ($t_s$) is 7 nm for the entire structure. The thickness of the silicon nitride cladding for the modulator ($h_m$) is slightly larger than the waveguide (set to 315 nm) to account for the lower index of the AZO layer as opposed to air. The AZO layer ($t_m$) is set at 350 nm thick.

The proposed modulator structure is based from a hybrid plasmonic waveguide design using TiN and silicon nitride, shown to achieve low-loss propagation at 1.55 μm [55]. By using a waveguide as the base structure, the insertion loss into the device is minimized, allowing for many devices to be cascaded while maintaining appreciable throughput. A schematic of the modulator, coupled to an input and output waveguide, is illustrated in Fig. 5. A plasmonic wave is injected into the structure from the left and a control pulse is used to modify the permittivity of the as-grown AZO sample from the top. This results in two simultaneous effects: 1) The as-grown AZO has a stronger absorption and 2) The careful balance of the effective index is perturbed. Increased absorption clearly leads to a reduction of the signal, but during modulation there is also a mismatch in the modes supported within the two structures which leads a to larger insertion loss. This is of course desired in the on-state (i.e. high loss state), but decreases the performance in the off-state.

As a first-order optimization of the structure, the silicon nitride thickness $h_w$ should be selected to minimize propagation losses in the waveguide and the thickness $h_m$ should be selected such that the wavevector of the mode in the modulator is close to that in the waveguide (see Supplementary Information). For an operating wavelength of 1.3 μm, the entire modulator structure was simulated in 2-D from the side so that propagation through both the connecting waveguides and the modulator stack could be analyzed. The optical pumping was simulated by simply altering the carrier concentration of the as-grown AZO film in a manner which corresponds to the values extracted from the pump-probe experiments (an incident fluence of 0 and 3.9 mJ/cm² was assumed). A 7 nm thick TiN strip with permittivity $\varepsilon = -49 + i16$ (as obtained from spectroscopic ellipsometry, see Supplementary Information) is used with a sapphire (n = 1.75) substrate and silicon nitride (n = 2) cladding. Vertical cut lines of the mode inside the waveguide Fig. 6(a) and modulator 6(b) are depicted for the on and off-state as well as the signal power as a function of the propagation distance in Fig. 6(c) (modulator length is highlighted in red). Although subtle due to the selection of the modulator's geometry, the mode in the off‑state is plasmonic, as an exponential decay in the substrate (z < 0) is evident near the metal (z = 0). Note that the profile in the silicon nitride (z > 0) is always Gaussian, resulting in the hybrid plasmonic-photonic nature of the waveguide.[55, 56] However, once pumped, the delicate index balance is no longer present and the mode exhibits an almost fully Gaussian profile (Fig. 6(b) on-state). This is largely due to the fact that the plasmonic mode is no longer supported in the biased modulator structure and the light has scattered to an available photonic mode. As a result of this and the increase in the absorption of the as-grown AZO, a signal modulation depth of 0.4 dB/μm is achieved. The total insertion losses in the structure are 0.4 dB for the 6 μm section shown here, or 0.06 dB/μm, which is the result of a low off-state propagation loss and an optimal mode overlap between the waveguide and modulator. In addition, further optimization of the structure, namely the initial permittivity of the as-grown AZO layer, can lead to increased performance which is expected to approach an extinction ratio of 1 dB/μm without sacrificing insertion loss or speed.

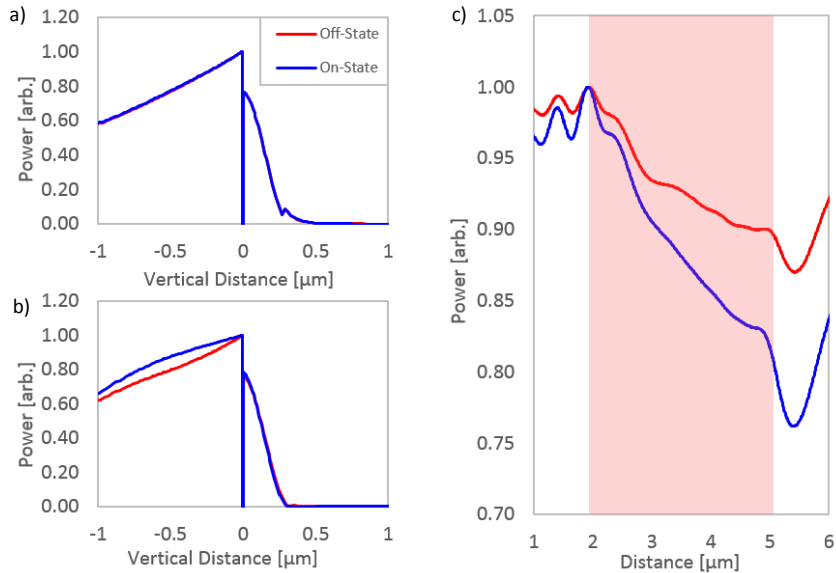

Fig. 6. a) Vertical mode cutline in the waveguide for the on and off-states. These curves overlap. b) Vertical mode cutline in the modulator for the on and off-states, illustrating the shift from a plasmonic mode to a photonic mode (i.e. not supported). c) Power flow in the structure along the direction of propagation. The modulator section is highlighted in red.

Unlike many other all-optical modulators, the design presented here is shown to be very simple yet effective, achieving speeds consistent with even the fastest nonlinear all-optical modulators (and orders of magnitude faster than other methods [42, 45]) while maintaining a respectable modulation depth for minimal fluence levels. In addition, this device uses practical materials which can be integrated into existing CMOS manufacturing technologies, as well as an eminently practical geometry, requiring just two standard photolithography steps (one for the waveguide and one for the as-grown AZO layer). Consequently, this alternative plasmonic modulator, powered by the tunability of as-grown AZO, represents a viable device for high-performance all-optical circuits.

## 5. Conclusion

The ability to achieve large modulation without sacrificing speed is critical for future photonic and hybrid photonic-electronic devices. Transparent conducting oxides, and aluminum doped zinc oxide in particular, are very promising materials for realizing high‑performance dynamic devices and can be modified either with the application of an electric field or through optical excitation. Although both techniques are favorable methods for realizing dynamic devices, all-optical methods are beneficial due to high operational speeds and the lack of RC‑delay limits. Here, a unique fabrication strategy is introduced to optimize the photodynamic properties of aluminum doped zinc oxide thin films while achieving ENZ properties at telecommunication wavelengths. Both the reflectivity and transmissivity were investigated using ultrafast pump-probe spectroscopy, whereby a large ($\Delta R/R_o$ = 40% with $\Delta T/T_o$ = 30%) and ultrafast (< 1 ps) response in both transmission and reflection was observed for all applied fluences. By modeling the physics behind the carrier dynamics, the excess carrier density within the material was extracted along with the recombination time of the process, which is believed to be the result of defect enhanced recombination, enabling verified speeds several orders of magnitude faster than traditional semiconductor materials. With only a few micro Joules of applied energy, the average carrier concentration within the bulk of the material (~350 nm) was changed by roughly 10%, resulting in a raw modification of the carrier density which is 10× larger than reported electrical biasing of TCOs. Providing a more uniform distribution of carriers within the bulk of a material, without sacrificing speed, is beneficial for a wide range of applications such as optical modulators and tunable metasurfaces to more efficiently utilize the entire available dynamic material. In addition, an all-optical plasmonic modulator with THz bandwidth using a suite of practical materials was proposed which demonstrated the potential of the as-grown AZO in a device under realistic experimental conditions. The modulation depth within the structure was shown to be 0.4 dB/μm with an extremely low insertion loss < 0.1 dB/μm, enabling the construction of high-performance, large scale nanophotonic circuits without the need for amplifiers. Consequently, AZO is shown to be a unique, promising, and CMOS-compatible material for high‑performance dynamic devices in the near future.

**Funding Information**

This research was supported by the following grants ONR-MURI grant N00014-10-1-0942, NSF MRSEC Grant DMR1120923), and AFOSR Grant FA9550-14-1-0138. M. Ferrera would like to acknowledge support from the Marie Curie Outgoing International Fellowship contract no. 329346.

**Acknowledgment**

We thank Urcan Guler and Amr Shaltout for useful discussions during this work.

# Supplementary Materials

1. **Oxygen-Deprived AZO Deposition and Characterization**

Aluminum-doped ZnO (AZO) films were deposited by pulsed laser deposition (PVD Products Inc.) using a KrF excimer laser (Lambda Physik GmbH) operating at a wavelength of 248 nm for source material ablation. A 2wt% doped AZO target was purchased from the Kurt J. Lesker Corp. with a purity of 99.99% or higher [1]. The energy density of the laser beam at the target surface was maintained at 1.5 J/cm2 and deposition temperature was 75°C. We maintained the oxygen pressure under 0.01 mTorr to achieve additional free carrier from the oxygen vacancies. The prepared thin films were characterized by spectroscopic ellipsometry (J. A. Woollam Co. Inc.) in the spectral region from 300 - 2500 nm. The dielectric function of the films was retrieved by fitting a Drude (when $\omega_o = 0$) and Lorentz oscillator model (Eq. S1) to the ellipsometry data. To probe the electrical properties of thin films such as mobility and carrier concentration, we carried out the hall measurement (MMR Technologies) at the room temperature.

$$\varepsilon = \varepsilon_\infty + \sum \frac{\omega_p^2}{(\omega_o^2 - \omega^2) - i\omega\Gamma} \qquad (S1)$$

2. **TiN Deposition and Characterization**

TiN is deposited using reactive magnetron sputtering in a nitrogen/argon environment (PVD Products Inc.). The substrate temperature is set to 800°C during deposition to obtain more metallic properties [2]. The linear optical properties of TiN used in modeling the plasmonic modulator are derived from spectroscopic ellipsometry (J. A. Woollam Co. Inc.) where the thickness is determined from TEM (see Fig. S1(a)). A Drude + Lorentz model (Eq. S1) with one Lorentz oscillator and two Drude terms (i.e. $\omega_o = 0$) is used to fit the data [2]. The extracted linear optical properties of the 7 nm thick TiN layer are shown in Fig. S1(b).

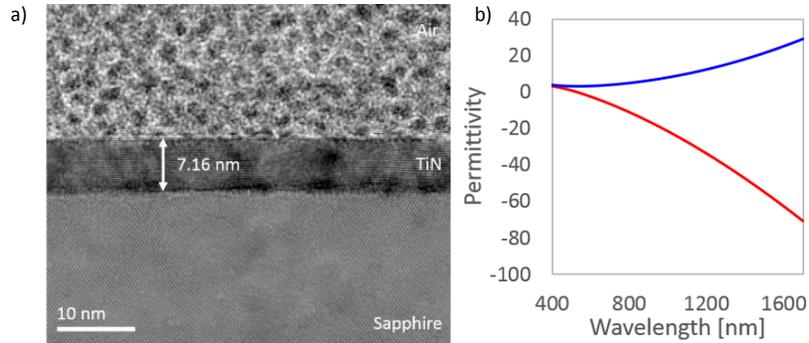

Fig. S1. a) TEM image of 7 nm thick TiN layer on c-sapphire. The cubic structure of TiN is evident due to the epitaxial growth on sapphire. b) Linear optical properties of TiN as extracted from spectroscopic ellipsometry measurements.

3. **Experimental Setup and Methods**

An amplitude Ti:sapphire laser system (Spectra Physics) was used as the excitation source for this experiment, producing pulses of ~120 fs at 1 kHz and 800 nm. The output was then fed through an optical parametric amplifier (TOPAS Inc.) and set to an output of 650 nm. In our system, 650 nm is generated as the second harmonic (SH) of 1300 nm, and therefore both the 1300 nm and 650 nm beams exit the system collinearly. These beams were then guided to our experiment and separated using a dielectric mirror with high reflectivity above 1060 nm. The 1300 nm beam is first filtered to remove any residual 650 nm light and guided through a retroreflector placed on a computer controlled stage, providing a delay line. To generate the pump beam, the 650 nm beam was focused into a BBO crystal properly oriented to phase match the SH. Any residual 650 nm was removed with a short pass filter. Both beams were then focused onto the sample such that the pump:probe beam area is roughly 4:1. The probe beam was collected in both transmission and reflection by a biased InGaAs photodiode connected to a lock-in amplifier which is locked to the laser repetition rate. The signal at the lock-in amplifier, as well as the z-position of the stage, was collected by custom software.

## 4. Theoretical Modeling

Using the linear properties of the AZO and fused silica substrate, the absorption of the transmitted (i.e. $T_{AZO} = 1 - R_{AZO}$) pump photon fluence per unit distance into the material is calculated, which corresponds to the excess carrier density. The incident flux was considered to have a symmetric a two dimensional Gaussian profile with $1/e^2$ width as determined from measurements. The induced excess carrier density at every point on the surface of the sample was then averaged over the thickness of the material and treated as a uniform effective medium which varied in one dimension (i.e. parallel to the surface due to pump beam shape). The temporal response of the system was modeled through the well-known dynamics of pump-probe experiments [3]:

$$f(t) = A_1 \exp\left(\frac{-t}{\tau_1}\right)\left[1 - erf\left(\frac{w}{2\tau_1} - \frac{t}{w}\right)\right] \quad \textbf{(S2)}$$

where $A_1$ is the amplitude of the process defined by the generation of excess carriers, $\tau_1$ is the recombination time of the excess carriers, and $w$ is cross-correlation width of the pump and probe pulses related the full-width at half-maximum of the Gaussian pulse by $w = 1/\sqrt{2\ln 2}\,\tau_{FWHM}$. This can be further extended to account for multiple decay processes by summing several iterations of Eq. (1) with varying amplitudes and decay times. For our purposes only one term was required. Once the z-averaged density of the excited carriers was determined for the initial pumping case, the entire distribution was multiplied by Eq. (1) to determine the change in the carrier concentration in time. Fig. S2 illustrates a 2-D plot of the z-averaged carrier density in time where the y axis is the position along the surface of the sample and the x-axis is time. The peak in excess carrier density here corresponds to point in time of the maximal change in the reflection and transmission as shown in Fig. S2.

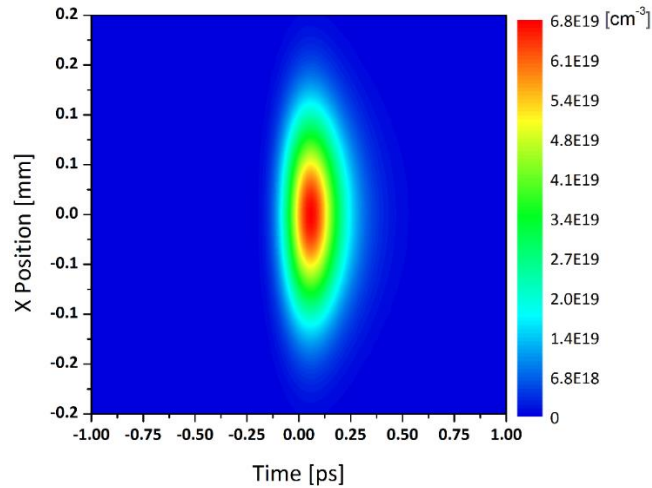

Fig. S2. 2-D plot illustrating the excess carrier distribution in spatial coordinate x (parallel to the sample's surface) and in time for the 350 nm AZO sample under a fluence of 3.9 mJ/cm$^2$. The maximum induced carrier density corresponds to the maximal change in the reflection and transmission.

Using the Drude model, which captures the optical response of the quasi-free excited carriers, the change in the optical properties was determined by adding the contribution of the excess electrons to the un-pumped (intrinsic) permittivity. The loss term in the Drude model was calculated based on mobility in the film as determined from Hall measurements. Following, the transfer matrix approach was used to determine the linear reflection and transmission of the multilayer sample for both the pumped and un-pumped case as a function of the x-position and time. The results are reported as the percent change in relation to the un-pumped case which was taken as $\Delta R/R_o$, and likewise for transmission. Subsequently, the percent change in the reflection and transmission was multiplied by the probe beam Gaussian profile to account for the spatial overlap of the two beams, and averaged over the $1/e^2$ width of the probe to better approximate the measured signal. Finally, the change in the reflection and transmission was determined in time.

The calculated change in reflection and transmission was matched to experiments through a series of steps. First, the pulse width of our laser is ~120 fs which dictates the rise time of the process. Due to experimental error in recording the absolute z-position of the computer controlled stage (i.e. delay time), the entire experimental trace is then shifted in time to overlap with the rise time of the laser fitted curve – taking special consideration to overlap points for t < 0 where the response is almost entirely dictated by the pump probe cross-correlation. Secondly, the decay time of the transient is adjusted to match the exponential decay of the experiment, focusing solely on the reflection measurement. The amplitude of the transient is determined by the fluence used during experiment and was only slightly adjusted at the end of the fitting to improve the results. All adjustments made to the incident fluence are within the experimental error, generally < 10% modification.

## 5. Plasmonic Modulator Optimization

A first order optimization of the plasmonic modulator structure was completed by ensuring an ideal mode overlap between the waveguide and modulator structures. This was achieved using eigenmode analysis in COMSOL Multiphysics by sweeping the thickness of the cladding and observing propagation constants. The waveguide propagation loss versus silicon nitride thickness for a 7 nm thick TiN strip is shown Fig. S3 along with a mode profile. For thicknesses less than 280 nm (shaded in red in Fig. S3), the mode was photonic in nature, having its peak in the substrate and not near the metal's surface. These modes are not considered in this analysis, and correspondingly a silicon nitride thickness of 285 nm was chosen for the waveguide. This mode had a wavevector $k_{spp} = 8.4579 \times 10^6$ - 74.44i (a propagation length more than 6 mm for a mode size of 3.5 μm). The same analysis was repeated for the modulator stack with a 350 nm layer of AZO (base permittivity $\varepsilon = 0.22 + i0.26$) on top of the silicon nitride, and the height $h_m$ was chosen to most closely match the k of the waveguide mode – a thickness of 315 nm. This ensures that in the off-state, the plasmonic mode is supported in both structures and minimizes the insertion loss.

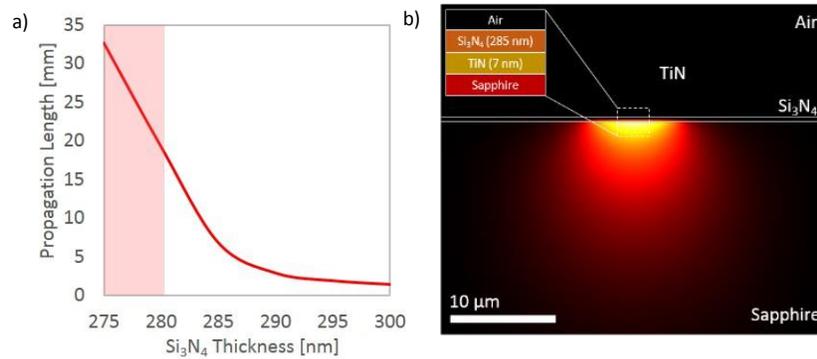

Fig. S3. a) Propagation length of the waveguide plasmonic mode versus the silicon nitride cladding thickness. The shaded region are modes with a photonic nature and are not considered in this analysis. b) Mode profile of the hybrid plasmonic mode in the waveguide for a silicon nitride thickness of 285